\documentclass[english]{elsart1p}
\usepackage{epsfig,amsmath,amssymb,graphics,graphicx,color,pstricks,natbib} 
\def\uy{\underline{y}}
\def\ux{\underline{x}}
\def\uxi{\underline{\xi}}
\def\ind{{\mathbb I}}
\begin{document}
\begin{frontmatter}

\title{ Constraint satisfaction problems and neural networks:
a statistical physics perspective
} 
\author{Marc M\'ezard}
\address{ LPTMS, UMR 8626 CNRS et Univ. Paris-Sud, 91405 Orsay CEDEX, France}
\author{Thierry Mora}
\address{Lewis-Sigler Institute for Integrative Genomics, Princeton University, Princeton, NJ 08544, USA}
\begin{abstract}

A new field of research is rapidly expanding at the crossroad between
statistical physics, information theory and combinatorial optimization. In
particular, the use of cutting edge statistical physics concepts and methods
allow one to solve very large constraint satisfaction problems like random
satisfiability, coloring, or error correction.

Several aspects of these developments should be relevant for the understanding
of functional complexity in neural networks. On the one hand the message
passing procedures which are used in these new algorithms are based on local
exchange of information, and succeed in solving some of the hardest
computational problems. On the other hand some crucial inference problems
 in neurobiology, like those generated in multi-electrode recordings,
naturally translate into hard constraint satisfaction problems.

This paper gives a non-technical introduction to this field, emphasizing the
main ideas at work in message passing strategies and their possible
relevance to neural networks modelling. It also introduces a new
message passing algorithm for inferring interactions between variables from
correlation data, which could be useful in the analysis of multi-electrode recording data.
\end{abstract}
\date{\today}

\end{frontmatter}

\section{Introduction: Constraint Satisfaction Problems}
Engineers offer encounter problems with many degrees of freedom (`variables')
but also many constraints. The problem is to find a value of the variables
which satisfies all constraints, or the most probable configuration of
variable given the constraints and some a priori measure. Obvious applications
are scheduling (classes, airplanes...), or job assignment. But similar
problems occur in various branches of scientific activity, and are crucial in
several domains. To be short we shall focus here on four of them. The
satisfiability problem is at the core of the theory of computational
complexity in computer science. Error correcting codes are one of the main
topic of information theory. Learning from examples is a basic process in
cognitive neuroscience. Reconstruction of neuron interactions from
multi-electrode recording is a problem which is becoming more and more
important.

All these problems can be formulated in a common language \citep{MM_prep}, and have a strong
relationship to fundamental issues in statistical physics like the existence
of phase transition, and the possibility of glassy phases. They can also be
cast into a somewhat generic formalism, based a graphical representation of
the topology of constraints \citep{KscFreLoe}, which allows to apply a general `message passing'
strategy to all of them. Some of these message passing algorithms have
actually shown strikingly good performance, solving some problems in
satisfiability or perceptron learning that are unreachable by any other
algorithms. It is interesting in itself to understand how fundamental issues
in computational complexity and information processing can be formulated in
the same language as relevant problems in neuroscience, the main aim of this
paper is to give some clues on these connexions.
\section{Satisfiability}
The problem of satisfiability involves $N$ Boolean variables ${ x_i
 \in \{T,F\} }$. There exist thus { $2^N$} possible configurations of
 these variables. The constraints take the special form of `clauses',
 which are logical `OR' functions of the variables. For instance the
 clause $ x_1 \vee x_{2} \vee \bar x_3$ is satisfied whenever $x_1=T$
 or $x_{2}=T$ or $x_3=F$ (the bar means negation: $\bar T=F$ and $\bar F=T$). Therefore, among the $8$ possible
 configurations of $ x_1, x_{2}, x_3$, the only one which is
 forbidden by this clause is $x_1=x_{27}=0$, $x_3=1$. An instance of
 the satisfiability problem is given by the list of all the clauses it contains.
The problem is to find a choice of the Boolean variables (called an 'assignment')
such that all constraints are satisfied. When there exists such a
choice the corresponding instance is said to be `SAT', otherwise it is
`UNSAT', and one typically seeks a configuration of variables which violates the smallest number of constraints.

Satisfiability plays an essential role in the theory of computational
complexity, because many other difficult problems like the traveling
salesman, the colouring of graphs, scheduling, protein folding, can be mapped
`polynomially' to it. It was the first problem which has been shown to be
`NP-complete' \citep{Cook}. This means that if one could find an algorithm that
solves satisfiability in a `polynomial' time (growing like a power of $N$),
one could also solve all these other problems in polynomial time: life would
be much easier, in particular the life of scientists... This is generally
considered unlikely, but the corresponding mathematical problem (whether the
NP class is distinct or not from the `P' class of problems which are solvable
in polynomial time) is an important open problem in mathematics.

\begin{figure}
\begin{center}
\includegraphics[angle=0,width=0.4\columnwidth]{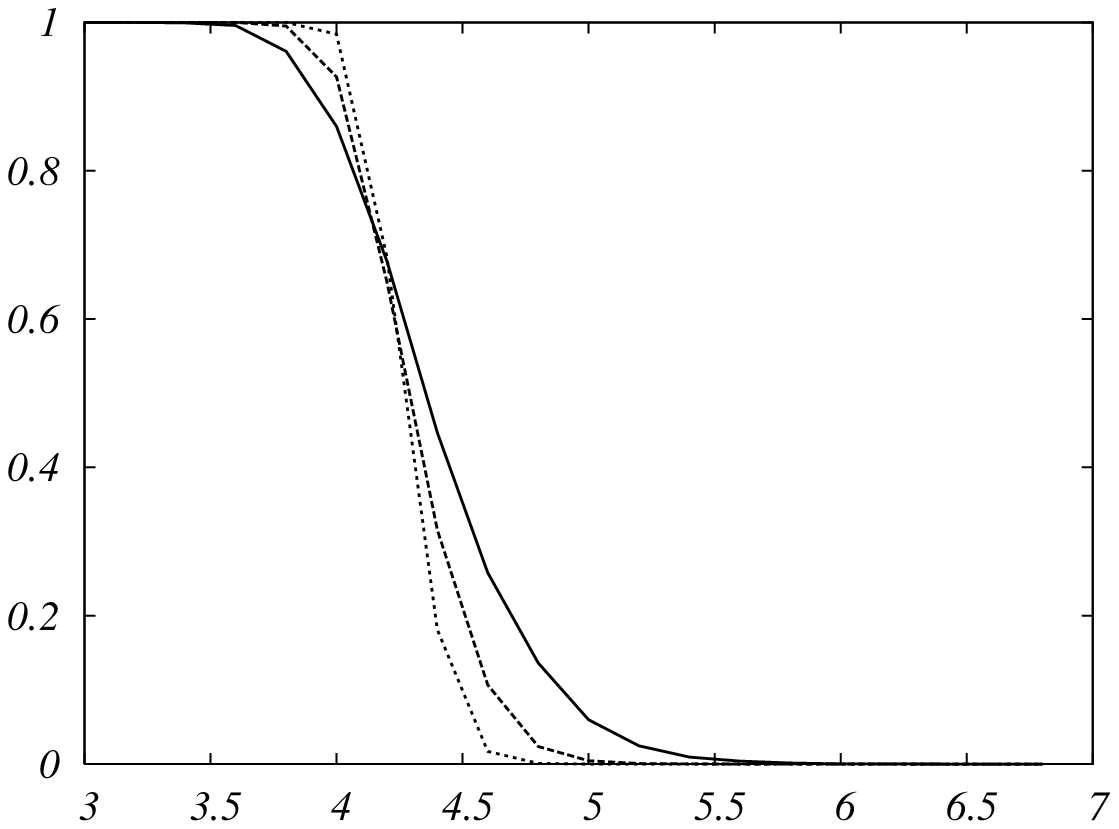}
\hspace{0.5cm}
\includegraphics[angle=0,width=0.4\columnwidth]{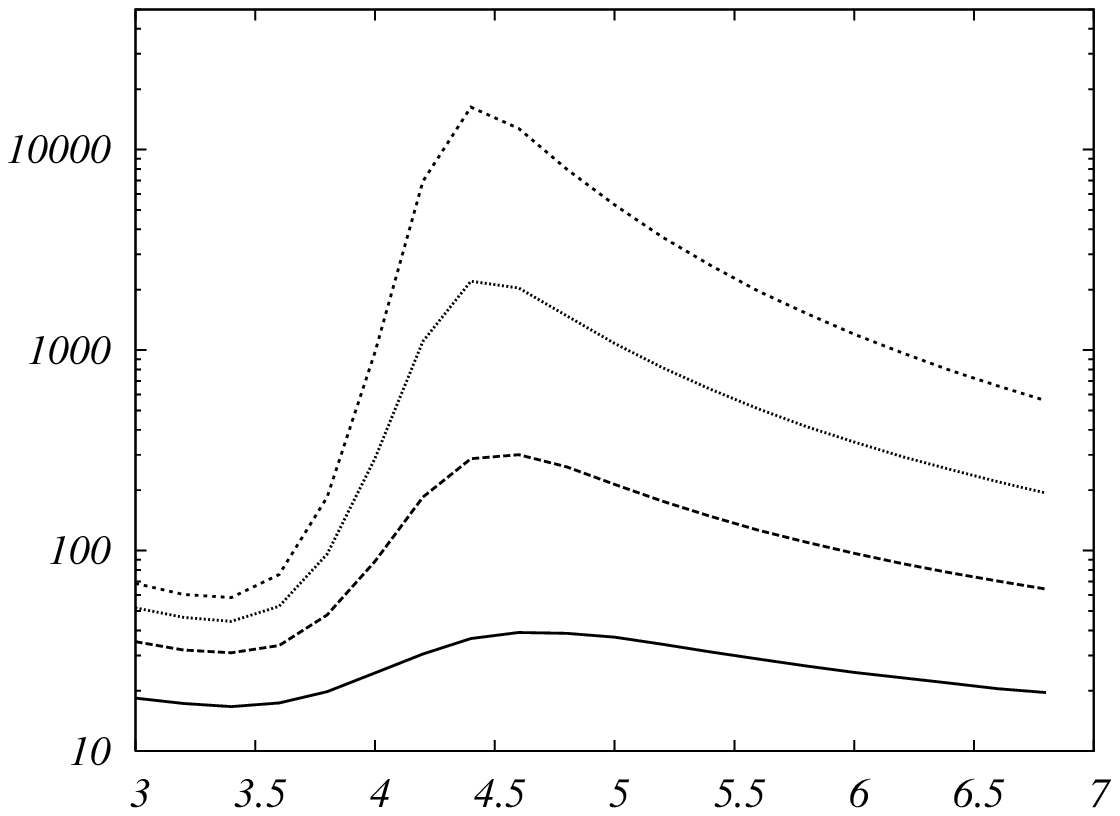}
\put(-80,-7){$\alpha$}
\put(-255,-7){$\alpha$}
\put(-160,55){$P_N$}
\put(-335,55){$P_N$}
\caption{Left: Probability that a formula generated from the random $3$-SAT
ensemble is satisfied, plotted versus the clause density $\alpha$.
The curves  correspond to $N=50$ (full line), $N=100$ (dashed), $N=200$
(dotted). The transition between satisfiable and unsatisfiable formulas becomes 
sharper as $N$ increases.
Right:
Computational effort.  Plotted is the computer time (in arbitrary units) required to
find a solution, or prove that there is no solution, versus the
clause density $\alpha$.  From bottom to top: $N=50$, $100$, $150$, $200$.}
\label{fig:alphac_SAT_num}
\end{center}
\end{figure}

The result of Cook is a worst case analysis of the satisfiability problem.
However it appears more and more important to study {\bf{`typical case'}}
complexity of satisfiability problems by introducing some classes of
instances. A much studied class is the random `3-SAT' problem. Each clause
contains exactly three variables chosen randomly in $\{ x_1,..,x_N\}$, and
each variable is negated randomly with probability $1/2$. This problem is
particularly interesting because its difficulty can be tuned by varying one
single control parameter, the ratio $ \alpha=\frac{M}{N}$ of constraints per
variable. One expects intuitively that for small $\alpha$ most instances are
SAT, while for large $\alpha$ most of them are UNSAT. Numerical experiments
have confirmed this scenario, but they indicate actually a more interesting
behavior. The probability that an instance is SAT exhibits a sharp crossover,
from a value close to $1$ to a value close to 0, at a threshold $\alpha_c$
which is around $4.3$. When the number of variables $N$ increases, the
crossover becomes sharper and sharper \citep{KirkSel1,KirkSel2}, as shown in
Fig.\ref{fig:alphac_SAT_num}. It has been shown that it becomes a staircase
behavior at large $N$ \citep{Friedgut}: almost all instances are SAT for
$\alpha<\alpha_c$, almost all instances are UNSAT for $\alpha>\alpha_c$. This
threshold behavior is nothing but a phase transition as one finds in physics,
and has been analyzed using the methods of statistical physics
\citep{KirkSel1,MZKST,MezParZec_science}.

A very interesting observation illustrated in Fig.\ref{fig:alphac_SAT_num}
is that the algorithmic difficulty of
the problem, measured by the time taken by the algorithm to answer if
a typical instance is satisfiable, also depends strongly on $\alpha$:
the  problem is easy when $\alpha $ is well below or well above
$\alpha_c$, and is much harder when $\alpha$ is close to
$\alpha_c$. Therefore the region of phase transition is also the
region which is difficult from the computational point of view.
\section{Error correction}
\begin{figure}
\begin{center}
\includegraphics[angle=0,width=0.4\columnwidth]{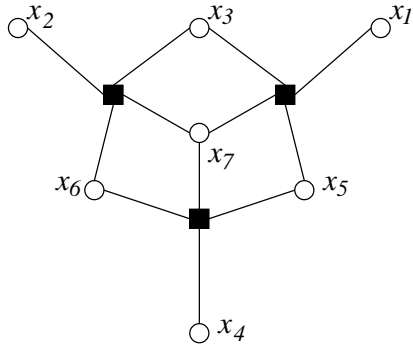}
\caption{Tanner graph representation of a parity check code. Here there are 7 bits related by three parity check equations.
Each square represents a parity check: it enforces the constraints that the sum of bits connected to it must be even}
\label{fig:tannerHamming}
\end{center}
\end{figure}

One of the fundamental problems in information theory consists in correcting transmission errors that always occur when a message is sent through a communication channel \citep{RiU05,MonUrbHouches}. This is done by adding redundancy. In codes based on parity constraints,
the message which is sent is chosen in a pool of `codewords'. A codeword is a set of $N$ bits $x_1,\cdots,x_N$,
where $x_i\in\{0,1\}$, which satisfies $M$ parity check equations taking the form:
\begin{equation}
x_{i_1(a)}+\cdots + x_{i_K(a)}= \text{even}
\end{equation}
For each $a\in\{1,\cdots,M\}$ there is one such equation, characterized by the set of bits $i_1(a),\cdots,i_K(a)$
which are involved in it. So the codebook, i.e. the set of codewords, is the set of solutions to these $M$ constraints. It is conveniently 
represented graphically as in Fig.~\ref{fig:tannerHamming}. Because the code is based on a system of linear equations, if they are designed to be independent, which is usually the case, the number of codewords will be $2^{N-M}$: the code transmits $N-M$ effective  bits of information, the extra $M$ bits are used to introduce redundancy and possibly correct errors.

How does one correct errors? Imagine for simplicity that a codeword $\ux= x_1,\cdots,x_N$ is sent through a `binary symmetric channel', which flips each bit independently with probability $p<1/2$. The received message is $ \uy= y_1,\cdots,y_N$,
where $y_i=x_i$ with probability $1-p$, and $y_i=1-x_i$ with probability $p$. Decoding means trying to infer the sent codeword
$\ux$ given the received one $\uy$. For this we write the probability that the sent message  was a set
of bits $\ux'= x_1',\cdots,x_N'$:
\begin{equation}
P(\ux'|\uy)=\frac{1}{Z} \prod_i\left[(1-p)\delta_{x_i',y_i}+p \delta_{x_i',1-y_i}\right] 
\prod_{a=1}^M \ind\left( x_{i_1(a)}'+\cdots + x_{i_K(a)}'= \text{even} \right)
\end{equation}
where the first terms come from our knowledge of the channel, and the last ones enforces the fact that the sent codeword is known to satisfy the parity check equations ($\ind(A)$ is an indicator function equal to one if the statement $A$ is true, equal to $0$ if it is not true). Decoding amounts to finding the most probable codeword given the received 
message, i.e. finding the set of bits $\ux'$ which maximizes $P(\ux'|\uy)$. This is in general another difficult, NP-complete, problem. But we will see that it can done efficiently with a message passing procedure called Belief Propagation (BP) if the noise level $p$ is not too large.

Low Density Parity Check (LDPC) codes are based on random constructions in which the parity check equations are generated randomly \citep{GallagerThesis}. For instance in
regular $(l,k)$ codes one generates equations such that each equation contains $k$ variables, and each variable appears in $l$ equations. In the large code limit $N\to\infty$ one finds two phase transitions when one varies $p$. The first one is the threshold for decoding through BP: it works almost always when $p<p_d$, it fails almost always if $p>p_d$. The second 
one is the threshold for decoding through exact inference (computing the true maximum of $P(\ux'|\uy)$). It works almost always when $p<p_c$, it fails almost always if $p>p_c$. For instance in a $(l=3,k=6)$ regular LDPC codes, the two thresholds
are $p_d= 0.084$ and $p_c=0.101$, while Shannon's theorem states that perfect decoding should be possible
up to $p=0.110 $, and impossible above. In practice the relevant threshold is $p_d$. This is because BP decoding is fast (it typically takes a time that grows linearly with $N$), while exact inference is much too slow (its time grows exponentially with $N$). Optimized LDPC codes can have a threshold $p_d$ which gets quite close to the Shannon limit \citep{RiU05,MonUrbHouches}.
\section{Two problems in neuroscience}
\subsection{Supervised learning}

Learning and memory tasks are believed to occur in neural systems through changes of synaptic strengths. Despite years of efforts, the precise way these changes are implemented in the brain for specific tasks is poorly understood. In the scenario of supervised learning, synaptic changes are monitored by a feedback signal carrying information about the success of the intended task. The perceptron classification problem is the prototypical example of supervised learning: given a set of training patterns ($\uxi^1,\ldots,\uxi^M$), where each $\uxi^a$ is a vector of $N$ binary variables ($\xi^a_i=\pm 1$, $i=1,\ldots,N$), we want to learn the correct synaptic weights $w_i$ leading to the classification of these inputs into two classes, $C_+$ and $C_-$, using a feed-forward network called {\em perceptron}:
\begin{equation}\label{eq:perceptron}
\textrm{for each }a=1,\ldots,M,\qquad\mathrm{sign}\left(\sum_{i=1}^Nw_i\xi_i^a\right) = \sigma_a
\end{equation}
where we require that $\sigma_a=+1$ if $\uxi^a$ belongs to class $C_+$, and $\sigma_a=-1$ if $\uxi^a$ belongs to class $C_-$.

\begin{figure}
\resizebox{\linewidth}{!}{\input{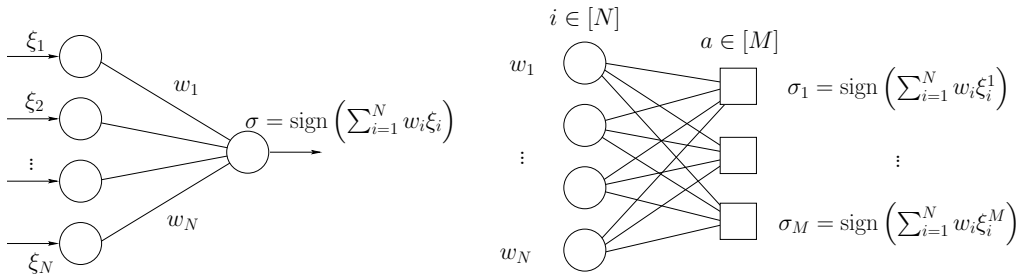}}
\caption{\label{fig:perceptron}Left: a perceptron is a feed-forward network that takes a pattern $\uxi$ as an input, and outputs a binary variable $\sigma$. Right: training of the perceptron viewed as a constraint satisfaction problem (factor graph representation, see further). Weights are variables (circles), and each pattern to be classified defines a constraint (squares).}

\end{figure}

Interestingly, this problem can be formulated as a constraint satisfaction problem, whose graph representation is given by the right panel of Fig.~\ref{fig:perceptron}. The weights $w_i$ are the unknown variables, and each pattern defines a constraint through Eq.~\eqref{eq:perceptron}. 

Efficient algorithms for solving this problem are known in the case of analog synaptic stengths (real $w_i$) \citep{Rosenblatt62}. However, recent experimental studies have shown that some synapses undergo changes in the form of jumps between a finite number of stable states \citep{PetersenMalenka98,ConnorWittenberg05}. Unfortunately, this discreteness makes the classification problem much harder: for instance, the task of learning binary weights $w_i=\pm 1$ is NP-complete \citep{BlumRivest92}. Although it has been known for years that a perceptron with binary synapses can in principle be trained to classify up to $M=\alpha_c N$ random patterns in the limit of large $N$, with $\alpha_c\approx 0.83$ \citep{KrauthMezard89b}, until recently no algorithm was known that could even perform this task for an extensive number of patterns (i.e. $M=\alpha N$ with $N\to\infty$ and $\alpha$ fixed), emphasizing the difficulty of the problem.

Like for error-correcting codes, message passing procedures provide a viable solution to this hard problem. The learning task can be handled approximately by algorithms derived from Belief Propagation \citep{BraunsteinZecchina06}. Somewhat surprisingly, these techniques perform well for large random problems, even relatively close to the theoretical threshold $M/N=\alpha_c$. An on-line, biologically relevant variant of BP, which can still classify an extensive number of patterns, has also been showcased as a plausible learning mechanism for realistic neural networks \citep{BaldassiBraunstein07}.

\subsection{Inferring neuronal couplings from multielectrode recordings}
\label{sec:maxent}

Recent experimental studies indicate that correlations play an important role in the retinal code \citep{SchneidmanBerry06}. In these experiments, many cells from a retinal ganglion patch are recorded simultaneously by a dense electrode array.
It was shown that individual cells do not carry independent pieces of information, but rather respond cooperatively through effective pairwise interactions. This suggests that the stimulus is represented in a redundant manner reminiscent of error-correcting codes. We will see that the problem of learning effective pairwise interactions between neurons from the observed data can also be formulated in our common statistical physics language.

Formally, the neural response of a retinal patch can be binned and represented by a string of binary variables. For each time bin of size $\delta t$ (with e.g. $\delta t=20$ ms), labelled by $t$, the neural response is coded by a binary word $\ux^t$, where $x^t_i=+1$ if neuron $i$ has fired in that time bin, and $x^t_i=-1$ otherwise. The neuronal response is stochastic in nature and can be described by a probability distribution $P(\ux)$, which accounts for both stimulus and noise fluctuations. Beside its interest for itself, a correct estimation of $P(\ux)$ is also important for the brain, as it may be used downstream the retina to evaluate the likelihood of spiking events, which in turn can be used to detect `abnormal' stimuli, or to perform classification tasks.

In the limit of a large integration time $T$, the probability distribution can in principle be measured through direct sampling:
\begin{equation}
P(\ux)\approx \frac{1}{T}\sum_{t=1}^T \delta_{\ux,\ux^t}
\end{equation}
In practice however, neither we nor the brain itself can handle such a large amount of data. If $N\approx 200$ is the number of cells in a patch, the number of pattern probabilities to be stored is $2^N\approx 10^{60}$, much more than any realistic integration time or storage capacity. One must thus recourse to simplifying assumptions. The simplest one is the independent approximation, which formally corresponds to factorizing the probability: $P(\ux)=\prod_{i=1}^N (1+x_im_i)/2$. One then just needs to measure the average $m_i:=\langle x_i\rangle$ of each neuron activity in order to reconstruct the full probability distribution  (brackets denote expectations with respect to $P(\ux)$). Unfortunately, this approximation fails to correctly render some important statistical properties of the collective response, including the law governing the total number of spikes in the population. This prompts us to take into account the correlative structure of the response.

The first step beyond independence is to consider pairwise correlation functions:
\begin{equation}
\chi_{ij}=\langle x_ix_j\rangle -\langle x_i\rangle\langle x_j\rangle,
\end{equation}
These numbers measure the propensity of pairs of neurons to spike cooperatively rather than independently. An approximate probability distribution, that reproduces these correlations as well as the average firing probabilities $(1+m_i)/2$ with minimal constraints, can be constructed using the principle of maximum entropy \citep{Jaynes49,SchneidmanStill03}. We look for a distribution $P^{(2)}(\ux)$ of maximum entropy
\begin{equation}
S:=-\sum_{\ux}P^{(2)}(\ux)\log P^{(2)}(\ux)
\end{equation}
that matches the one and two-point correlation functions of the observed response:
\begin{equation}
\chi^{(2)}_{ij}=\chi_{ij},\qquad m^{(2)}_i=m_i.
\end{equation}
This distribution, which is uniquely defined, has been shown to account for most (90\%) of the correlative structure of as many as 40 neurons recorded silmutaneously in the retina \citep{SchneidmanBerry06}.

With the help of Lagrange multipliers one can show that the Maximum Entropy distribution takes the form:
\begin{equation}\label{eq:Ising}
P^{(2)}(\ux)=\frac{1}{Z}\exp\left(\sum_{i} h_i x_i+\sum_{i>j} J_{ij}x_ix_j\right)
\end{equation}
where $Z$ is a normalization constant. In physics terms this is a disordered Ising model. Usually, physicists face the problem of solving {\em direct} Ising problems, which typically consist in infering thermodynamical quantities, as well as magnetizations $m_i$ and correlation functions $\chi_{ij}$, from the external fields $h_i$ and couplings $J_{ij}$. This problem is computationally very hard in general, and there exist no simple relation between $(h_i,J_{ij})$ on the one hand, and $(m_i,\chi_{ij})$ on the other: an exact estimate requires summing over the $2^N$ possible configurations $\ux$.
Here we have to deal with the {\em inverse} Ising problem (inferring the couplings from the correlation functions), which is even harder.

This learning problem and its variants have become increasingly important recently. Besides its relevance to neural decoding, it is also useful for thinking about inference in protein interaction networks \citep{Tkacik07}, the correlative structure of some catalytic proteins \citep{SocolichLockless05,RussLowery05}, and even the statistical properties of four-letters words in English \citep{StephensBialek07}.

A number of algorithmic strategies, mostly based on Monte-Carlo sampling, have been proposed to learn the couplings from the correlation functions \citep{AckleyHinton85,BroderickDudik07}. Very little is known, however, about possible neural implementations of this learning task. We will see that strategies based on message-passing ideas may provide leads on that question.

\section{The message passing strategy}
All the problems we have seen so far can be formulated in a common language. We have $N$ variables $(x_1,\cdots,x_N)$,
taking value in some space $X$, and they are linked by constraints of probabilistic nature: each constraint $\psi_a$
links  the variables with labels $i_1(a),\cdots,i_K(a)$, in the form of a probabilistic factor
$\psi_a(x_{i_1(a)},\cdots , x_{i_K(a)})$. In the case of hard constraints like parity checks the hard constraint takes
value $1$ if the check is satisfied, $0$ otherwise. In other cases it can take intermediate values, like for instance the
factors $\left[(1-p)\delta_{x_i',y_i}+p \delta_{x_i',1-y_i}\right]$ due to the received message in coding. The problem is defined by a probability
distribution 
\begin{equation}
P(\ux)=\frac{1}{Z}\prod_a \psi_a(x_{i_1(a)},\cdots , x_{i_K(a)})
\label{eq:pdef}
\end{equation}
Our goal is twofold. On the one hand we want to study the properties of one given instance: compute the
marginal distributions $P(x_i)$, or find the $\ux$ which maximizes $P(\ux)$. On the other hand when $P$ is generated from an ensemble which allows to consider the large $N$ limit one would like to understand the phase diagram of the problem, like the thresholds $p_d$ and $p_c$ that we defined in decoding.

\begin{figure}
\begin{center}
\epsfig{file=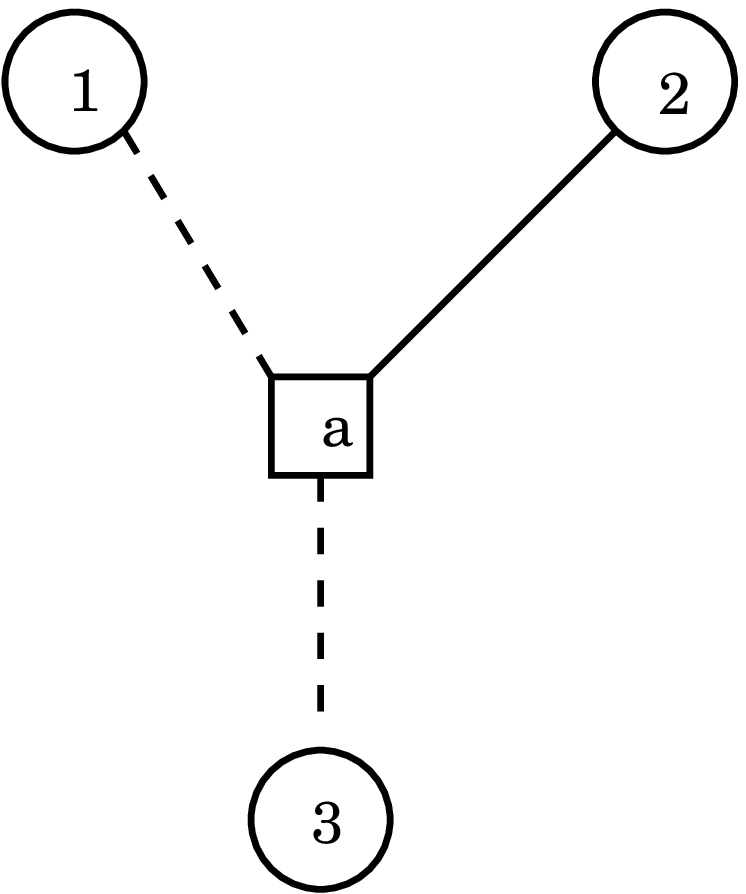,width=0.2\textwidth}
\hspace {3 cm}
\epsfig{file=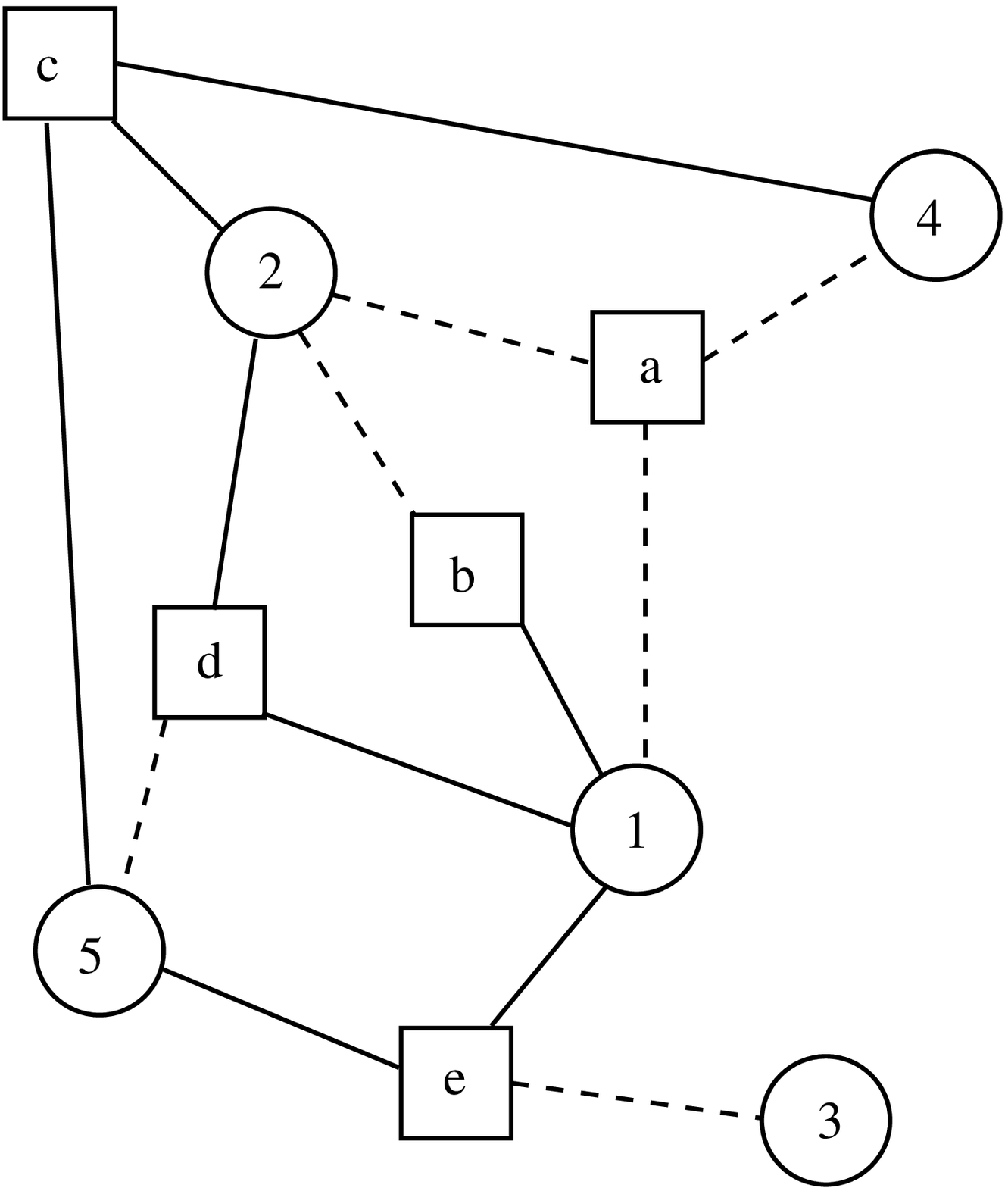,width=0.35\textwidth}
\end{center}
\caption{Factor graph representation of satisfiability: A variable is represented by
a circle. A constraint is represented by a square, connected with a full (resp. dashed)
 line to a variable when this variable appears as such (resp. negated) in the clause.
Left hand side: The clause $ \bar x_1 \vee x_2 \vee \bar x_3$. Right hand side:
the factor graph representing the formula: 
${( \bar x_1 \vee  \bar x_{2} \vee \bar x_4) }{\wedge }{ ( x_{1} \vee 
\bar x_2)}{\wedge }{( x_2 \vee   x_{4} \vee  x_5) } {\wedge }{ ( x_{1} \vee 
 x_2 \vee \bar x_5)}{\wedge }{(  x_1 \vee \bar  x_{3} \vee  x_5) }$}
\label{fig_clause}
\end{figure}

Eq.~\eqref{eq:pdef} is not the most general probability distribution between $N$ variables: the crucial point is that each $\psi_a$ involves only a finite number of variables. When $N$ is very large, $P$ induces a topological structure in the space of variables that we shall exploit. The factor graph representation is a very convenient way of characterizing this structure. Each constraint $\psi_a$
is represented by a function node (square), connected to the various variables (circles) which
appear in the constraint \citep{KscFreLoe}. An example for satisfiability is described in Fig.\ref{fig_clause}.
The Tanner graph of code is nearly a factor graph: one just needs to add to it degree 1 function nodes connected to each variables, accounting for the factor  $\left[(1-p)\delta_{x_i',y_i}+p \delta_{x_i',1-y_i}\right]$. The factor graph of the perceptron learning problem is shown on Fig.~\ref{fig:perceptron}.

If the factor graph were a tree, it would be easy to solve our problem (for instance find marginals). The idea of BP is to write `mean field' like equations that would be exact on a tree, and try to use them also in more general (and more interesting) cases. BP equations are self-consistency relations between two types of `messages', $\eta_{i\to a}$ and $\eta_{a\to i}$. On trees, $\eta_{i\to a}$ can be interpreted as the probability measure on $x_i$ when the factor node $a$ has been removed, while $\eta_{a\to i}$ is the probability measure on $x_i$ when all factors neighboring $i$, expect $a$, have been removed. Denoting $\partial a=\{i_1(a),\cdots,i_K(a)\}$ the neighborhood of $i$, and $\partial i$ the neighborhood of $a$, BP equations read \citep{MM_prep}:
\begin{eqnarray}
\eta_{a\to i}(x_i)&=&\frac{1}{z_{a\to i}}\sum_{\ux_{\partial a\backslash i}}\psi_a(x_{i_1(a)},\ldots,x_{i_K(a)})\prod_{j\in \partial a\backslash i} \eta_{j\to a}(x_j)\label{eq:bp1}\\
\eta_{i\to a}(x_i)&=&\frac{1}{z_{i\to a}}\prod_{b\in \partial i\backslash a}\eta_{b\to i}(x_i)\label{eq:bp2}
\end{eqnarray}
where the $z$'s are normalization constants.
In practice, these equations are solved by iteration (with parallel or random update schedules) until a fixed point is reached. Convergence is typically met in linear time. This makes BP a very fast algorithm.
At the fixed point, the probability measure on $x_i$ is given by:
\begin{equation}\label{eq:bp3}
P_i(x_i)=\frac{1}{z_i}\prod_{a\in \partial i}\eta_{a\to i}(x_i)
\end{equation}
Thermodynamical quantities such as the free-energy $-\log Z$ can also be derived \citep{MM_prep} from the messages $(\eta_{i\to a},\eta_{a\to i})$.

Note that while convergence and accuracy are garanteed when the graph is a tree, BP equations sometimes fail to find the correct fixed point or provide a poor approximation of the probability measure when the graph is loopy. This can happen when there are many small loops, or when correlations build up across the graph. To overcome the first issue, generalized Belief Propagations (GBP) schemes have been proposed \citep{Yedidia01b}. The second issue, which is related to the partition of the measure $P$ into a multiplicity of disconnected `states', can be handled by an extension of BP called Survey Propagation (SP) \citep{MezZec_SAT,BraMezZec_SAT}.

As we mentionned earlier, BP is the best known solver for LDPC codes, provided that the channel noise is not too high.
While BP can also handle random satisfiability problems for small enough clause densities $\alpha$, SP becomes necessary as one gets to higher $\alpha$, where problems become hard. SP can find solutions to $3$-SAT instances for up to $10^7$ variables at $\alpha=4.25$, very close to the satisfiability threshold $\alpha_c$ \citep{MezZec_SAT}.

Beside their efficiency, the appeal of message passing procedures like BP resides in their local nature: information is propagated along the edges of the graph, and each message is updated using only other messages coming into the same node. This makes them highly amenable to parallelization. It is also tempting to make the connection with learning mechanisms in the brain, whereby synaptic strengths change only according to the activity of its neighboring neurons. And indeed, the engineering of BP/SP-inspired algorithms for the perceptron show that learning rules using only post and pre-synaptic activities, as well as error signals, suffice to implement efficient learning \citep{BaldassiBraunstein07}.

\section{An application: the inverse Ising problem}
We now study a novel application of message-passing to the inverse Ising problem introduced in Section \ref{sec:maxent}. As in the perceptron, the proposed method relies on local exchanges of information between variables.

\begin{figure}
\begin{center}
\resizebox{.5\linewidth}{!}{\input{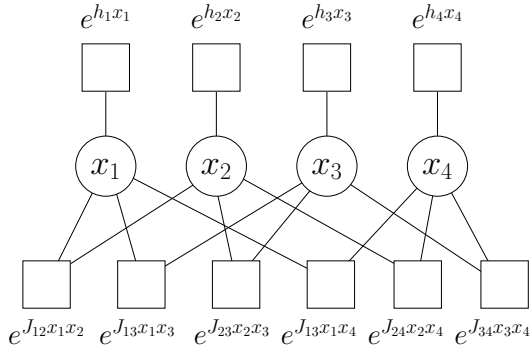}}
\caption{\label{fig:factorising}Factor graph representation of the Ising model Eq.~\eqref{eq:Ising}.}
\end{center}
\end{figure}

Let us start with the direct problem, whose factor graph is represented in Fig.~\ref{fig:factorising}. BP can be used to compute probability measures on single variables (i.e. local magnetizations $m_i$), but it does not give information on the two-point correlation functions $\chi_{ij}$. To access this information we will need to go a bit further. We shall make use of the fluctuation-dissipation relation, which offers a convenient way to estimate pairwise correlation functions using the derivatives of magnetizations:
\begin{equation}
\chi_{ij}=\chi_{ji}=\frac{\partial m_i}{\partial h_j}=\frac{\partial m_j}{\partial h_i}.
\end{equation}

But first we need to adapt the language of BP to the Ising model.
The binary nature of Ising variables allows us to reduce BP messages to single numbers:
\begin{equation}
\eta_{i\to a}(x_i)=\frac{1+x_i m_{i\to a}}{2},\qquad \eta_{a\to i}(x_i)=\frac{1+x_i m_{a\to i}}{2}
\end{equation}
These messages $m_{i\to a}$ and $m_{a\to i}$ are called `cavity' magnetizations, as they are defined on amputed graphs. Note that when factor $a$ is just a field contribution $e^{h_ix_i}$, the message is trivial. When factor $a$ is a interaction contribution $e^{J_{ij}x_ix_j}$, we rewrite for convenience $m_{i\to j}:=m_{i\to a}$.

The iteration of BP equations, along with Eq.~\eqref{eq:bp3}, allows to compute the $m_i$'s.
We now define a new type of messages, called cavity susceptibilities, and defined as:
\begin{equation}
\chi_{i\to j,k}:=\frac{\partial m_{i\to j}}{\partial h_k}
\end{equation}
These messages are tied by a new set of self-consistency equations, called `susceptibility propagation' equations, simply obtained as the derivatives of BP equations \eqref{eq:bp1}, \eqref{eq:bp2} with respect to $\{h_k\}$. They reflect how small local perturbations can propagate through the graph to remote variables, even when these variables and the perturbation are not directly linked. As in BP, these equations can be solved iteratively.
When convergence is reached, the total susceptibilities $\chi_{ij}$ are given by derivatives of Eq.~\eqref{eq:bp3} with respect to $\{h_k\}$.

This susceptibility propagation algorithm has the same advantages and downsides as BP. While being relatively fast, it relies on the assumption that the behaviour of the model is not far from that of a tree factor graph. This can be true if the graph is sparse and locally tree-like, or if the interactions are small enough.

Susceptibility propagation (approximately) solves the direct Ising problem $(h_i,J_{ij})\to (m_i, \chi_{ij})$. How can we use it to solve the inverse problem? The key is to realize that although susceptibility equations are self-consistency equations on the messages, they can also be viewed as self-consistency equations on the `inputs' $(h_i,J_{ij})$ by simply extracting them from the belief and susceptibility propagation equations. For instance, on can write the following update rule for $J_{ij}$,
\begin{equation}\label{eq:upJ}
\tanh J_{ij}= \frac{\chi_{ij}-m_{i\to j} m_{j\to i}}{1-\chi_{ij} 
m_{i\to j}m_{j\to i}}.
\end{equation}
The rest of the iteration equations remains essentially unchanged, with the notable difference that now $(m_i,\chi_{ij})$ are treated as constants, while $(h_i,J_{ij})$ become the unknown variables to be updated.

\begin{figure}
\begin{center}
\resizebox{.9\linewidth}{!}{
\input{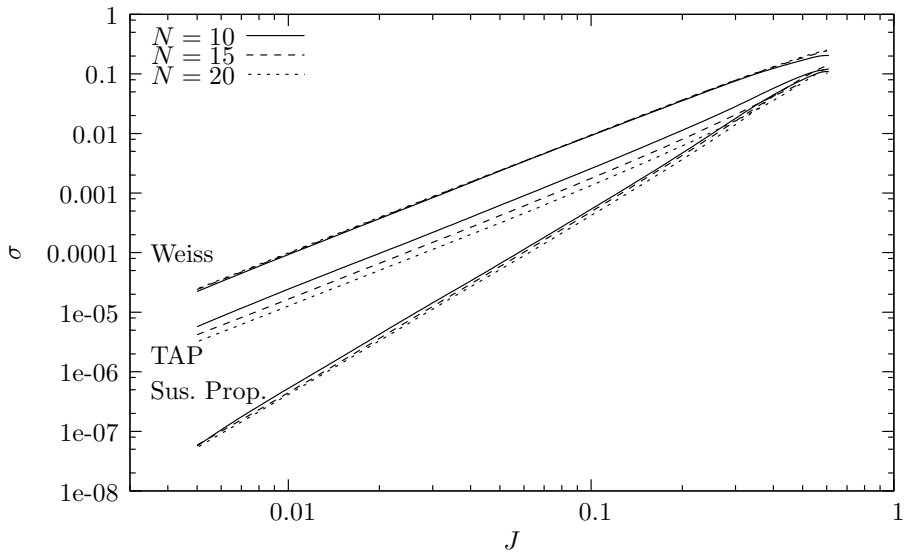}}
\caption{\label{fig:sk}Mean error $\sigma^2=\frac{N}{J^2}\langle( J'_{ij}-J_{ij})^2\rangle$ of the susceptibility propagation reconstruction algorithm presented in the text, compared against that of two mean-field schemes \citep{KappenRodriguez98} (Weiss: naive mean-field, TAP: mean-field with back-reaction term).}
\end{center}
\end{figure}

We have tested our algorithm on synthetic data. First we have considered a spin glass with random gaussian couplings $J_{ij}$ of zero mean and variance $J^2/N$, with no magnetic fields, $h_i=0$. This is the Sherrington-Kirkpatrick model. Small problems ($N=10,15,20$) are drawn at random and solved exactly by exhaustive enumeration. Then our algorithm tries to reconstruct the couplings $J_{ij}$ from the correlation functions. Its performance is shown on Fig.~\ref{fig:sk}, and is contrasted with other mean-field methods \citep{KappenRodriguez98}. Interestingly, all mean-field schemes fail for $J>1$, where the system notoriously becomes `glassy', with the onset of metastable states.

\begin{figure}
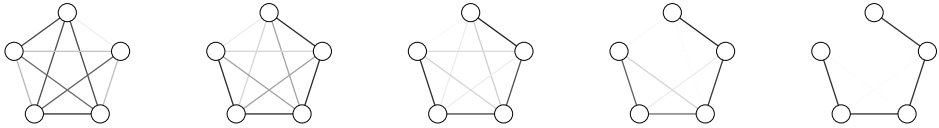

\begin{center}
\hskip -11cm
\begin{minipage}{.19\linewidth}
\resizebox{3\linewidth}{!}{\input{evol0}}
\end{minipage}
\begin{minipage}{.19\linewidth}
\resizebox{3\linewidth}{!}{\input{evol3}}
\end{minipage}
\begin{minipage}{.19\linewidth}
\resizebox{3\linewidth}{!}{\input{evol6}}
\end{minipage}
\begin{minipage}{.19\linewidth}
\resizebox{3\linewidth}{!}{\input{evol9}}
\end{minipage}
\begin{minipage}{.19\linewidth}
\resizebox{3\linewidth}{!}{\input{evol100}}
\end{minipage}
\caption{\label{fig:chain}Reconstruction of a small linear chain. Knowing only the correlation functions, and with no prior knowledge on the topology of the graph, the algorithm can infer both the structure and the numerical values of the interaction strengths. Here is shown the progress of the algorithm. The gray level of each edge codes for the couplings strength $J_{ij}$. The algorithm is started with random initial conditions (leftmost graph). Next are shown, from left to right, the couplings after 3, 6, 9 and 20 iterations.}
\end{center}
\end{figure}

Perhaps the power of susceptibility propagation is better shown on examples where it is supposed to be exact, namely, when the underlying topology is a tree. For simplicity we have tested our algorithm on linear chains. Provided that the couplings are not too large, we can reconstruct {\em both} the topology of the linear chain (i.e. the order of variables on the chain), and the exact strength of interactions between neighbours (see Fig.~\ref{fig:chain}). When the couplings are too large, the exact solution becomes unstable. This can partially be remedied, however, by making zero couplings more attractive in the equations, thus stabilizing sparse topologies.

A more systematic method for treating sparse networks is however needed. With
it, susceptibility propagation could be used as a comprehensive network
reconstruction algorithm, with possible applications to the inference of
Bayesian networks, Markov chains with arbitrary topologies, or in population
genetics.

\section{Conclusion}
The message passing strategy often provides the most efficient algorithms for
solving hard constraint satisfaction problems, or for inference in graphical
models. This is especially true when the factor graph representing the problem
has a local tree-like structure. It is particularly remarkable that some very
difficult problems, which cannot be solved by other methods, are solved by
procedures of local exchange of messages between the variables and
constraints. It is likely that recent developments in this domain can have
some impact in neuroscience, in at least two directions. First of all because
some major challenges in neuroscience, linked to the analysis of experimental
data, can themselves be formulated in terms of graphical or constraint
satisfaction problems. Secondly because the mere fact that distributed local
information exchange systems achieve this task is very appealing in the
perspective of information processing by the brain. 

\bibliographystyle{neuron}
\bibliography{tauc_ref}

\end{document}